\documentclass[pra,aps,onecolumn,showpacs,tightenlines]{revtex4}
\usepackage{graphics,bm}
\usepackage{graphicx}
\def\bmath#1{\mbox{\boldmath$#1$}}
\newcommand{\beq}{\begin{equation}}
\newcommand{\eeq}{\end{equation}}
\newcommand{\bqa}{\begin{eqnarray}}
\newcommand{\eqa}{\end{eqnarray}}
\def\gsim{\mathrel {\vcenter {\baselineskip 0pt \kern 0pt
\hbox{$>$} \kern 0pt \hbox{$\sim$} }}}

\protect

\begin{document}

\title{Vortex Dynamics Near the Surface of a Bose-Einstein Condensate}
\author{U. Al Khawaja}
\affiliation{ \it Physics Department, United Arab Emirates
University, P.O. Box 17551, Al-Ain, United Arab Emirates.}

\date{\today}

\begin{abstract}
The center-of-mass dynamics of a vortex in the surface region of a
Bose-Einstein condensate is investigated both analytically using a
variational calculation and numerically by solving the
time-dependent Gross-Pitaevskii equation. We find, in agreement
with previous works, that away from the Thomas-Fermi surface, the
vortex moves parallel to the surface of the condensate with a
constant velocity. We obtain an expression for this velocity in
terms of the distance of the vortex core from the Thomas-Fermi
surface that fits accurately with the numerical results. We find
also that, coupled to its motion parallel to the surface, the
vortex oscillates along the direction normal to the surface around
a minimum point of an effective potential.

\end{abstract}


%\begin{multicols}{2}
 \maketitle

\section{Introduction}
\label{introduction}

Following the experimental observation of vortices in
Bose-Einstein condensates
\cite{matt,and,mad,mad2,haj,ram,abo,abo2,hod}, vortex dynamics has
been a subject of intensive interest
\cite{gora,butt,castin,dalfovo,feder,fetter,fetter2,
mad3,anatoly,james1,james2,usamarecent,rembert}. Theoretical
studies have been performed on the center-of-mass motion of the
vortex core in the bulk of the condensate
\cite{fetter,gora,rembert}. References \cite{gora,rembert} have
studied the dynamics of the core taking into account the
dissipation from the thermal cloud. Dissipation causes the vortex
to spiral out of the condensate giving it a finite lifetime in
agreement with the experimental observations \cite{and}. Once they
reach the surface of the condensate, the vortices decay into
surface excitations. It is also believed that vortices enter the
condensate from the surface at the point when surface excitations
become unstable
\cite{james1,dalfovo3,isos,mur,usamarecent,tsu,gardiner}. Hence,
it is worth investigating the details of the dynamics of a vortex
in the surface region of the condensate.

E. Lund {\it et al.} \cite{emil} have pointed out that near the
surface of the condensate, the quadratic trapping potential can be
approximated by a potential that is linear in distances normal to
the surface. As a result of this approximation, the equilibrium
density profile will be also linear in distances normal to the
surface of the condensate except for distances very close to the
Thomas-Fermi surface. An immediate question then arises: How will
the vortex move in such an inhomogeneous density background? The
fact that the density is linear makes it possible to perform
analytical treatment to the problem provided that the size of the
core of the vortex is much less than the distance over which the
background density varies significantly. This question has been
addressed extensively by Anglin~\cite{james2}, where the author
used a boundary-layer theory to calculate both the density profile
and phase of the vortex, from which he obtained the equilibrium
velocity of the vortex. It turned out that the component of the
velocity in the direction normal to the surface of the condensate
vanishes, while the component parallel to the surface is a
constant that depends on the distance between the core of the
vortex and the surface. In this paper we address basically the
same problem but using a variational calculation. The main aim is
to demonstrate that the variational calculation performed here
generates the previous knowledge about the vortex motion near the
surface and captures the main features of it. We go one step ahead
of previous works by obtaining a rather more accurate expression
for the component of the vortex velocity parallel to the surface.
Furthermore, we show that, within few coherence lengths from the
Thomas-Fermi surface, the vortex oscillates along the direction
normal to the surface around a minimum point of an effective
potential. These oscillations are coupled to its motion parallel
to the surface and disappear when the vortex is away from the
surface.

In our variational calculation we use an ansatz wavefunction that
takes into account the equilibrium density profile and phase of
both the background and the vortex. The variational parameters are
the coordinates of the vortex core and the center-of-mass velocity
components normal to the surface of the condensate and parallel to
it. The equations of motion for these parameters are then derived
from a lagrangian that corresponds to the time-dependent
Gross-Pitaevskii equation.

Using a linear density profile for the background is an
approximation that breaks down at the Thomas-Fermi surface. Our
variational calculation is, thus, expected to be accurate only
away from the Thomas-Fermi surface. Furthermore, we neglect the
internal dynamics of the core of the vortex. This is also
justified away from the Thomas-Fermi surface where the size of the
core of the vortex is almost constant.

In the following, we start, in subsection~\ref{sub1}, by writing
down the ansatz wavefunction. Using this wavefunction, we
calculate the lagrangian from which we derive the equilibrium
properties in subsection~\ref{sub2} and the equations of motion
for the variational parameters in subsection~\ref{sub3}. In
Section~\ref{numsec}, we present the results of the numerical
simulation of the vortex motion, and compare them with those of
the variational calculation. In Section~\ref{conc} we summarize
our main conclusions.

\section{Variational Approach}
\label{sec1}
%We start in this section by calculating the
%lagrangian a vortex moving in a linear background density. This is
%performed using an appropriate trial wavefunction that accounts to
%the phase and density profile of the vortex. In subsection
%\label{sub1} we calculate the equilibrium coordinates and velocity
%of the vortex by minimizing this energy functional. In subsection
%\ref{sub2}, we calculate the lagrangian of the moving vortex to
%obtain the time dependence of the coordinates and velocity of the
%vortex.

\subsection{The Lagrangian}
\label{sub1} 
\begin{figure}
\begin{center}
\includegraphics[width=10.cm]{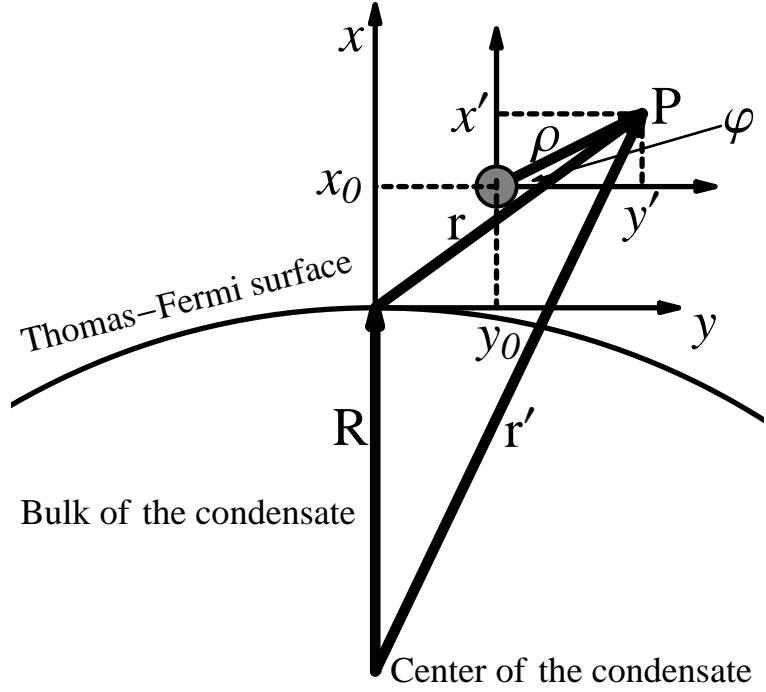}
\end{center}
\caption[a]{The planner geometry and coordinate system used in
this paper. The small circle represents the core of the vortex
which is of the order of the coherence length $\xi$. The
coordinates of the core of the vortex are $x_0$ and $y_0$. The
radius of the condensate is $R$. The bulk of the condensate is
located in the region $x<0$.  The point {\bf P} has cartesian
coordinates $x^\prime$ and $y^\prime$ related to the polar
coordinates $\rho$ and $\phi$ by $x^\prime=\rho\sin{\phi}$ and
$y^\prime=\rho\cos{\phi}$.} \label{fig1}
\end{figure}

The dynamics of the time-dependent Bose-Einstein
condensate is described by the Gross-Pitaevskii equation
\begin{equation}
\left[-{\hbar^2\over2m}\bmath{\nabla}^2_{\bf r^\prime}+V({\bf
r^\prime})+g|\psi({\bf r^\prime},t)|^2-\mu\right]\psi({\bf
r^\prime},t)=i\hbar{\partial\over\partial t}\psi({\bf
r^\prime},t)\;. \label{gp}
\end{equation}
Here $g$ is the effective two-particle interaction which is
proportional to the $s$-wave scattering length $a$ according to
$g=4\pi a\hbar^2/m$, where $m$ is the mass of an atom, and $\mu$
is the equilibrium chemical potential. The harmonic trapping
potential $V({\bf r^\prime})$ is given by
\begin{equation}
V({\bf r^\prime})={1\over2}m\omega_0^2{r^\prime}^2\;, \label{v}
\end{equation}
where $\omega_0$ is the characteristic frequency of the trap.
Here, we take for simplicity $V({\bf r^\prime})$ to be isotropic.
%,although this is not necessary in the limit which we take next.
For distances near the surface of the condensate, the quadratic
trapping potential $V({\bf r^\prime})$ can be approximated by a
linear potential \cite{emil,usama}
\begin{equation}
V({\bf r^\prime})\simeq V(R)+Fx \label{linearv}\;,
\end{equation}
where $x$ is a coordinate normal to the surface of the condensate,
$R$ is the Thomas-Fermi radius given by $V(R)=\mu$, and
$F=m\omega_0^2R$ is a force constant. The last equation is
obtained by using the transformation ${\bf r^\prime}={\bf r}+R{\bf
\hat{x} }$ in Eq.~(\ref{v}). This suggests a planner geometry in
which the surface of the condensate is approximated by an infinite
plane. The $x$-coordinate will be normal to the plane while the
$y$ and $z$ coordinates will be parallel to it. The origin of this
coordinate system is then shifted from the center of the
condensate to the surface. As a result of this shift, the bulk of
the condensate occupies the region $x<0$ and the surface region
will be near $x=0$. This is illustrated in Fig.~\ref{fig1}.

In this planner geometry, the Gross-Pitaevskii equation takes the
form
\begin{eqnarray}
\left[-{\hbar^2\over2m}\bmath{\nabla}^2_{\bf
r}+F\,x+g|\psi({\bf r
},t)|^2\right]\psi({\bf r},t)=i\hbar{\partial\over\partial t}
\psi({\bf r},t)\;.
\label{gp1}
\end{eqnarray}
The equilibrium density profile $n_0(x)=|\psi(x)|^2$ is obtained
from the last equation by setting the time derivative to zero
\begin{equation}
\left[-{\hbar^2\over2m}{d^2\over dx^2}+F\,x+g\,n_0(x)\right]\sqrt{n_0(x)}
=0\;.
\label{gp2}
\end{equation}
For large negative values of $x$, the kinetic energy operator can
be neglected~\cite{emil,baym} leading to the Thomas-Fermi
equilibrium density $n_0^{\rm TF}(x)=-(F/g)x$.

We consider a vortex with a core located at ${\bf
r}=x_0(t){\bf\hat x}+y_0(t){\bf \hat y}$ and axis along the
$z$-direction as shown in Fig.~\ref{fig1}. The vortex has a
center-of-mass velocity ${\bf \hat v}=v_x(t){\bf \hat
x}+v_y(t){\bf \hat y}$. Our trial wavefunction is given by
\begin{equation}
\psi({\bf r},t)=\sqrt{-Fx\over
g}\,\chi(\rho)\,\exp{(i\Phi(\rho,\phi))}\,
\exp{\left[i(m/\hbar)(xv_x+yv_y)\right]}\;. \label{psitrial}
\end{equation}
Here, the first factor on the right hand side is $\sqrt{n_0^{\rm
TF}(x)}$. The second factor corresponds to the square root of the
density of a vortex in a uniform background. The function
$\Phi(\rho,\phi)$ in the exponential is the phase of the vortex
that takes into account the inhomogeneouty of the background
density. The last factor represents the center-of-mass velocity of
the vortex. The polar coordinates
$\rho=\sqrt{(x-x_0(t))^2+(y-y_0(t))^2}$ and
$\phi=\tan^{-1}{\left((x-x_0)/(y-y_0)\right)}$ have the core of
the vortex as their origin, as indicated in Fig.~\ref{fig1}. Near
the surface, the Thomas-Fermi approximation breaks down and the
density background is nonlinear. The kinetic energy starts to
become important leading to an exponential decay of the density.
Our trial wavefunction does not take this nonlinearity into
account. Therefore, the results of this variational calculation
are not expected to be accurate in the surface region where the
Thomas-Fermi approximation breaks down.
%The effect of the nonlinear region near $x=0$ is accounted for in
%the next section where we solve numerically the time-dependent
%Gross-Pitaevskii equation, Eq.~(\ref{gp}).
%In addition, we believe it is interesting enough to address
%the question of vortex dynamics in a linear density background.

For the phase of the vortex we use an expression similar to the
one derived by Anglin \cite{james2}
\begin{equation}
\Phi(\rho,\phi)=\phi+{\rho\over 2x_0}\cos{(\phi)}\ln{\alpha
\rho\over x_0} \label{jphase}.
\end{equation}
The first term of this expression represents the usual
$\phi$-dependance of a vortex in a uniform background. The second
term results from the linear inhomogeneouty of the background. It
should be noted that the angle $\phi$ defined in this paper is the
complement of that defined in Ref. \cite{james2}. This expression
is obtained using a boundary-layer theory which is described here
briefly. Vortices have structure in their density over a small
length scale of the order of the coherence length
$\xi_0=1/\sqrt{8\pi a n_0(x_0)}$, and phase that extends over the
whole system. Here, $\xi_0$ is the coherence length calculated
with the background density at $x_0$, and
$\delta=(\hbar/mF)^{1/3}$ is the surface depth \cite{emil}. In the
boundary-layer theory, the vortex problem is first solved exactly
in the {\it inner} region $\rho \sim \xi_0$ and the {\it outer}
region $\rho\gg\xi_0$. The {\it inner asymptotics} of the outer
solution and the {\it outer asymptotics} of the inner solution are
then obtained by expanding the outer and the inner solutions in
powers of two perturbation parameters characteristic to the two
regions. Finally, the two solutions are matched to determine the
solution in the {\it intermediate} region. The velocity and phase
of the vortex are obtained as a result of this matching procedure.
The phase, given by Eq.~(\ref{jphase}), corresponds to the inner
asymptote of the outer solution. Thus, this expression is valid
only near the vortex and is not expected to be accurate for $\rho
\gg\xi_0$. However, in the present calculation, we take this
expression to be valid for all values of $\rho$. We account for
the fact that it is not valid for distances away from the vortex
core, by introducing in the argument of the logarithm the
parameter $\alpha$, which works as an effective cutoff on the
logarithmic term. The value of $\alpha$ is then determined by
fitting the velocity calculated variationally in
subsection~\ref{sub2} with the one obtained from the numerical
solution of the time-dependent Gross-Pitaevskii equation in
section~\ref{numsec}.

The vortex density profile is taken as the square of
\cite{fetter_old}
\begin{equation}
\chi(\rho)={\rho\over\sqrt{\rho^2+2\xi^2}}\;. \label{chi2}
\end{equation}
It turns out in subsection~\ref{sub3} that the equations of motion
of $y_0(t)$ and $v_y(t)$ are independent of $\chi$. Therefore,
specifying the functional form of $\chi$ is not necessary for the
$y$-component of the motion.

Due to the nonlocal nature of the vortex excitation, the energy of
the vortex diverges logarithmically with the size of the system
\cite{book}. To be able to perform the variational calculation for
a finite and constant number of atoms $N$, we consider in
calculating the lagrangian only the atoms within a cylinder of
radius $b$ and length $l$ such that the cylinder axis coincides
with the vortex axis. The fact that we restrict the number of
atoms within the cylinder to be constant, is guaranteed by
requiring $\psi({\bf r},t)$ to be normalized to $N$. The
normalized wavefunction takes the form
\begin{equation}
\psi({\bf r},t)=\sqrt{N\over lN_1}\sqrt{x\over
x_0(t)}\,\chi(\rho)\,
\exp{\left[i\left(\Phi(\rho,\phi)+(m/\hbar)(xv_x+yv_y)\right)\right]}\;,
\label{psitrial_norm}
\end{equation}
where
\begin{equation}
N_1=\int_0^b d\rho\,\rho\int_0^{2\pi}d\phi\,\chi^2\,. \label{n1}
\end{equation}
The lagrangian that corresponds to Eq.~(\ref{gp1}) is given by
\begin{equation}
L[\psi,\psi^*]= \int{d{\bf r}{i\hbar\over2}\left( \psi^*({\bf
r},t){\partial\psi({\bf r},t)\over\partial t} -\psi({\bf
r},t){\partial\psi^*({\bf r},t) \over\partial t} \right)-\Delta
E[\psi,\psi^*]}\,, \label{lagrangian}
\end{equation}
where $\Delta E[\psi,\psi^*]$ is the energy associated with the
presence of the vortex. This energy is obtained by subtracting the
energy of the {\it vortex-free} background from the vortex energy
\cite{book}, namely
\begin{equation} \Delta
E[\psi,\psi^*]=E[\psi,\psi^*]-E_{\rm vf}[\psi,\psi^*]~,
\label{defunc}
\end{equation}
where
\begin{eqnarray}
E[\psi,\psi^*]=\int_0^bd\rho\rho\int_0^{2\pi}d\phi\left[{\hbar^2\over2m}\left|\bmath{\nabla}\psi
({\bf r },t)\right|^2+F\,x|\psi({\bf r},t)|^2+{1\over2}g|\psi({\bf
r },t)|^4\right]\;, \label{energy}
\end{eqnarray}
and the vortex-free energy is
\begin{eqnarray}
E_{\rm vf}&=& \int_0^bd\rho\rho\int_0^{2\pi}d\phi\left[ {1\over
2}g\left({Nx\over\pi b^2 lx_0}\right)^2 +F\left({Nx\over \pi
b^2lx_0}\right)x \right]\,.
\end{eqnarray}
The terms on the right hand side of the last equation represent
the mean-field and trapping potential energies of the vortex-free
background, respectively. These were calculated using the
wavefunction
\begin{equation}
\psi_{\rm vf}(x)=\sqrt{N\over\pi b^2l }\sqrt{x\over x_0}
\label{vfwave}~.
\end{equation}
The prefactor of $\psi_{\rm vf}$ guarantees its normalization to
$N$. For the subtraction of the background energy from the vortex
energy to be correct, it is essential that the number of atoms of
both systems to be the same.

Using the trial wavefunction Eq.~(\ref{psitrial_norm}) in the
lagrangian of Eq.~(\ref{lagrangian}), we show in
Appendix~\ref{app} that the lagrangian per atom takes the form
\begin{eqnarray}
L/N&=&-(N_2(x_0)+N_3){\hbar^2\over2m}
-m(x_0\dot{v}_x+y_0\dot{v}_y)-N_5{m\dot{v}_x\over2x_0}\nonumber\\
&+&(1-2N_4(x_0)){\hbar\over
4x_0}(v_y-\dot{y}_0)-{1\over2}m(v_x^2+v_y^2)\nonumber\\
&+&N_4(x_0){\hbar^2\over4mx_0^2} -F\left({N_5\over2}-{
b^2\over4}\right){1\over x_0}\nonumber\\
&-&{1\over2}\gamma\left[N_6-{N_1\over\pi
b^2}+\left({N_7\over2}-{N_1\over4\pi}\right){1\over
x_0^2}\right]\label{lagapp2}~,
\end{eqnarray}
where the coefficients $N_1,\dots,N_7$ are functions of $b$, and
$N_2$ and $N_4$ have also $x_0$ dependence ( See
Appendix~\ref{app}.). The dot on $y_0$, $v_x$, and $v_y$ denotes
derivative with respect to time. The parameter $\gamma=gN/N_1l$ is
an energy that characterizes the average mean-field energy since
$N/N_1l$ represents the average density within the cylinderical
volume surrounding the vortex. This lagrangian will be the basis
of the calculations of the equilibrium and nonequilibrium
properties in the next two subsections.

\subsection{Equilibrium Properties}
\label{sub2} The equilibrium values of the variational parameters
can be obtained by minimizing the energy functional with respect
to the variational parameters.  The energy functional can be
readily obtained from Eq.~(\ref{lagapp2}) by setting time
derivatives to zero and then multiplying by -1, namely
\begin{eqnarray}
\Delta E[x_0,v_x,v_y]/N&=&
(N_2(x_0)+N_3){\hbar^2\over2m}+{1\over2}m(v_x^2+v_y^2)-(1-2N_4(x_0)){\hbar
v_y\over4
x_0 }\nonumber\\
&-&N_4(x_0){\hbar^2\over4mx_0^2} +F\left({N_5\over2}-{
b^2\over4}\right){1\over x_0}\nonumber\\
&+&{1\over2}\gamma\left[N_6-{N_1\over\pi
b^2}+\left({N_7\over2}-{N_1\over4\pi}\right){1\over
x_0^2}\right]\label{defunc2}~.
\end{eqnarray}
Minimizing $\Delta E$ with respect to $v_x$ gives $v_x=0$. As we
shall see in the next subsection, $v_x$ is proportional to
$\dot{x}_0$ in the limit $x_0\ll-\delta$. Therefore, $x_0$ will be
a constant throughout the motion of the vortex. In other words,
the vortex will be moving parallel to the surface of the
condensate in agreement with previous works~\cite{james2,fetter2}
and the results of the numerical simulation of
section~\ref{numsec}. Minimizing the energy with respect to $v_y$
yields
\begin{equation}
v_y(t)=\left(1-2N_4(x_0)\right){\hbar\over 4m x_0(t)}\,.
\label{vy1}
\end{equation}
In the hydrodynamic limit, $b\gg\xi$, the coefficient $N_4(x_0)$
is expanded in powers of $\xi/b$ as $N_4= -1/2+\ln{(\alpha
b/x_0)}+O\left((\xi/b)^2\ln{(b/\xi)}\right)$. Using this
expression in the last equation, it takes the form
\begin{equation}
v_y(t)={\hbar\over 2m x_0(t)}\left(1+\ln{x_0\over\alpha b
}\right)\,. \label{vy}
\end{equation}
The quantity ${\hbar/ 2m x_0}$ is the equilibrium velocity of a
vortex in a homogeneous density background located at a distance
$x_0$ from a hard wall and is moving parallel to it. The
logarithmic term corresponds, therefore, to the contribution of
the linear inhomogeneouty in the background density. The value of
$\alpha b$ is determined by fitting this expression to the
numerical values of $v_y(t)$. This is performed in section
\ref{numsec} where it turns out that $\alpha b\approx0.11$. Thus
Eq.~(\ref{vy}) reads
\begin{equation}
v_y(t)={\hbar\over 2m
x_0(t)}\left(3.2+\ln{x_0\over\delta}\right)\,. \label{vy2}
\end{equation}
This is to be compared with the expression derived by Anglin
\cite{james2} using a boundary-layer theory, namely
$v_y(t)={(\hbar/ 2m
x_0(t))}\left(1.96+(3/2)\ln{(x_0/\delta)}\right)$. In
Fig.~\ref{fig2}, we plot $v_y$ as a function of $x_0$ using
Eq.~(\ref{vy2}), the expression of Anglin, and the numerical
solution of the Gross-Pitaevskii equation. This figure shows that
while the expression of Anglin is accurate for large values of
$x_0$, it deviates from the numerical values of $v_y$ for values
of $x_0$ of order $\delta$. On the other hand, Eq.~(\ref{vy2})
describes accurately $v_y$ for both large and small values of
$x_0$. Equation~(\ref{vy2}) represents one of the main results of
this paper as it shows that the variational calculation does
indeed lead to a rather accurate description of the vortex
velocity parallel to the surface of the condensate.

Substituting the equilibrium expression of $v_y$ from
Eq.~(\ref{vy1}) in Eq.~(\ref{defunc2}), we obtain an energy
functional that is a function of $v_x$ and $x_0$, namely
\begin{eqnarray}
\Delta E[v_x,x_0]/N&=& {1\over2}mv_x^2 +U(x_0)\label{defunc3}~,
\end{eqnarray}
where
\begin{eqnarray}
U(x_0)&=& {\hbar^2\over2m}(N_2(x_0)+N_3)
+F\left({N_5\over2}-{b^2\over4}\right){1\over x_0}\nonumber\\
&+&{1\over2}\gamma\left[N_6-{N_1\over\pi
b^2}+\left({N_7\over2}-{N_1\over4\pi}\right){1\over
x_0^2}\right]\label{defunc4}~
\end{eqnarray}
is a function of $x_0$ that can be considered as an {\it effective
potential} for the vortex. Here, we have neglected a term
containing $N_4^2$ since it is one order of magnitude smaller than
the logarithmic term of Eq.~(\ref{jphase}). Inspection shows that
this effective potential has a minimum for negative values of
$x_0$. This means that the vortex can acquire oscillations around
this minimum. Taking this into consideration and the fact that for
large $x_0$, we have $v_x=\dot{x}_0$, suggests following a
collective coordinates approach to find the frequency of
oscillation \cite{usama}. In this approach the energy functional,
Eq.~(\ref{defunc3}), is put in the form of that of a simple
harmonic oscillator $\Delta E[v_x,x_0]=mv_x^2/2+k(x_0-x_{\rm
eq})^2/2$, where $x_{\rm eq}$ is the value of $x_0$ at which
$U(x_0)$ has a minimum. The frequency of oscillation will be
simply $\omega=\sqrt{k/m}$. This is achieved by expanding the
coefficients $N_1, N_2(x_0), N_5, N_6$, and $N_7$ in powers of
$\xi/b$ keeping terms up to the second order, namely:
%$N_1\simeq \pi
%b^2-2\pi\xi^2\ln{(b/\xi)}+O\left(\xi^2/b^2\right)$,
$N_2(x_0)=(1+\sqrt{1-(b/x_0)^2}\,)/2b^2+O\left((\xi/bx_0)^2\right)
$,
%$N_3\simeq2b^{-2}\ln{(b/\xi)}+ O\left((\xi/b)^{-3}\right)$,
%$N_4\simeq
%-1/2+\ln{(b/8x_0)}+O\left((\xi/b)^2\ln{(b/\xi)}\right)$,
$N_5= b^2/2+2\xi^2\ln{(b/\xi)}+8\xi^2(\xi/b)^2\ln^2{(b/\xi)}
+O\left((\xi/b)^4\ln^3{(b/\xi)}\right)$, $N_6=
1-4(\xi/b)^2\ln{(b/\xi)}+O\left((b/\xi)^3\right)$, $N_7=
b^2/2+2\xi^2\ln{(b/\xi)}+8\xi^2(\xi/b)^2\ln^2{(b/\xi)}
+O\left((\xi/b)^4\ln^3{(b/\xi)}\right)$~. In this limit, the
effective potential reduces to
\begin{eqnarray}
U(x_0)={\hbar^2\over mb^2}\ln{b\over\xi}
+\ln{b\over\xi}\left(1+{2\xi^2\over
b^2}\ln{b\over\xi}\right)\gamma{\xi^2\over x_0^2 }
+\ln{b\over\xi}\left(1+{2\xi^2\over
b^2}\ln{b\over\xi}\right)F{\xi^2\over x_0 }
 \label{ue22p}~,
\end{eqnarray}
which has a minimum at
\begin{equation}
x_{\rm eq}=-{2\gamma\over F}\left(1-{2\xi^2\over
b^2}\ln{b\over\xi}\right) \label{xequi}~.
\end{equation}
Expanding $U(x_0)$ around $x_{\rm eq}$, we obtain
\begin{eqnarray}
U(x_0)=-{F^2\xi^2\over 4\gamma}\ln{b\over\xi}+\left({F^4\xi^2\over
16\gamma^3}\ln{b\over\xi}\right)\left(x_0-x_{\rm eq}\right)^2
 \label{ue23p}~.
\end{eqnarray}
The energy functional becomes
\begin{eqnarray}
\Delta E[x_0,v_x]/N&=& -{F^2\xi^2\over
4\gamma}\ln{b\over\xi}+{1\over2}mv_x^2
+{1\over2}\left({F^4\xi^2\over
8\gamma^3}\ln{b\over\xi}\right)\left(x_0-x_{\rm
eq}\right)^2\label{defunc5}~.
\end{eqnarray}
The frequency of vortex oscillation is then given by
\begin{equation}
\omega=\sqrt{{F^4\xi^2\over 8m\gamma^3}\ln{b\over\xi}} \label{w}~.
\end{equation}
Noting that $\gamma=\hbar^2/2m\xi^2\sim n(x_0)$ and that in the
limit $|x_0|\gg\delta$, the Thomas-Fermi approximation gives
$n(x_0)\sim|x_0|$, we conclude that
$\omega\sim\sqrt{|x_0|^{-4}\ln{b/\xi}}$. Thus, for large $x_0$,
these vortex oscillations disappear and $x_0$ will be constant as
we have also found above by minimizing $\Delta E$ with respect to
$v_x$.

In the next subsection, we derive equations of motion that
describe the vortex motion and oscillations.

\subsection{Equations of Motion}
\label{sub3} The equations of motion follow from the
Euler-Lagrange equations of the lagrangian Eq.~(\ref{lagapp2}),
namely
\begin{equation}
\dot{v}_y-{d\over dt}\left[(1-2N_4(x_0)){\hbar\over4mx_0}\right]=0
\label{eqnofm1}~,
\end{equation}
\begin{equation}
\dot{y}_0-v_y+(1-2N_4(x_0)){\hbar\over4mx_0}=0 \label{eqnofm2}~,
\end{equation}
\begin{equation}
v_x-{d\over dt}\left(x_0+{N_5\over2x_0}\right)=0 \label{eqnofm3}~,
\end{equation}
\begin{eqnarray}
&&-\left(1-{N_5\over2x_0^2}\right){d^2\over dt^2
}\left(x_0+{N_5\over2x_0}\right) -N^\prime_2(x_0){\hbar^2\over2m}\nonumber\\
&+&{F\over m}\left({N_5\over2}-{b^2\over4}\right){1\over x_0^2}
+{\gamma\over m}\left({N_7\over2}-{N_1\over4\pi}\right){1\over
x_0^3}=0 \label{eqnofm4}~.
\end{eqnarray}
The prime on $N_2(x_0)$ denotes a derivative with respect to
$x_0$. In obtaining the last equation in terms of $x_0$ only, we
have used Eqs.~(\ref{eqnofm2}) and (\ref{eqnofm3}). In addition,
we have neglected terms containing $N_4^2(x_0)$ and
$N_4(x_0)N_4^\prime(x_0)$ since these terms are of an order of
magnitude less than the logarithmic term in the phase
$\Phi(\rho,\phi)$ of Eq.~(\ref{jphase}).

From the first two equations, we obtain quite generally
\begin{equation}
{\ddot y_0(t)}=0\,.
\label{ydd}
\end{equation}
This is a general result in the sense that it does not depend on
the functional form of $\chi$, the upper limit of the radial
integration $b$, the cutoff of the logarithmic part of the phase
$\alpha$, or $x_0$. The coordinates of the vortex are then
described completely by Eqs.~(\ref{eqnofm4}) and (\ref{ydd}).

In the limit $|x_0|\gg\delta$, Eqs.~(\ref{eqnofm3}) and
(\ref{eqnofm4}) reduce to $\dot{v}_x=x_0$ and $\ddot{x}_0=0$,
respectively. Therefore, $x_0$ will be in general a linear
function of $t$, namely $x_0(t)=c_1t+c_2$, where $c_1$ and $c_2$
are constants. We have shown in the previous subsection that
$v_x=0$ for the energy to be minimum. This results in that
$c_1=0$, and thus $x_0$ is a constant as we have also found in the
previous subsection. We notice also that, in this limit,
Eq.~(\ref{eqnofm2}) reduces to $v_y=\dot{y}_0$. This implies that
in the limit $|x_0|\gg\delta$, the variational parameters $v_x$
and $v_y$ are the conjugate velocities of $x_0$ and $y_0$,
respectively.

In the limit $b\gg\xi$, the coefficients $N_1,\dots,N_7$ can be
expanded in powers of $\xi/b$, as was shown in the previous
subsection, and Eq.~(\ref{eqnofm4}) simplifies to
\begin{eqnarray}
&&\left(1-{b^2\over4x_0^2}\right){d^2\over dt^2
}\left(x_0+{b^2\over4x_0}\right)+{dU(x_0)\over dx_0}=0
\label{eqnofm4s}~,
\end{eqnarray}
where $U(x_0)$ is the effective potential of the vortex defined in
Eq.~(\ref{ue22p}). Expanding the last equation around $x_{\rm eq}$
and using Eq.~(\ref{ue23p}), we get
\begin{eqnarray}
&&\left(1-{b^2\over4x_{\rm eq}^2}\right){d^2\over dt^2
}\left(x_0-x_{\rm eq}\right)+\omega(x_0-x_{\rm eq})=0
\label{eqnofm56s}~.
\end{eqnarray}
Using the initial conditions $x_0(0)=x_{\rm eq}+X_0$ and
$\dot{x}_0(0)=0$, the solution of the last equation reads
\begin{equation}
x_0(t)=x_{\rm eq}+X_0\cos{\Omega t} \label{sola}~,
\end{equation}
where $\Omega=\omega/(1-b^2/4x_{\rm eq}^2)$ and $\omega$ is given
by Eq.~(\ref{w}). Notice that in the limit $b\gg\delta$, the
quantity $b^2/x_{\rm eq}^2\rightarrow0$, which leads to
$\Omega\rightarrow\omega$. Equation~(\ref{eqnofm4}) can be solved
numerically for given values of $b$. We plot this numerical
solution in Fig.~\ref{fig3} together with analytical solution
Eq.~(\ref{sola}). This figure shows that the analytical solution
is a reasonable approximation for the exact solution of
Eq.~(\ref{eqnofm4}).

To ensure that these oscillations are not an artifact of our
variational calculation, we have solved numerically the
time-dependent Gross-Pitaevskii equation for initial vortex
distances from the Thomas-Fermis surface that are of order
$\delta$. We found that the vortex indeed oscillates in the
$x$-direction as predicted by the variational calculation.
Furthermore, we show next that we can obtain good agreement
between the variational calculation and the numerical one for both
the frequency and amplitude of the oscillations. In the present
variational calculation, the amplitude of vortex oscillation $X_0$
and frequency $\omega$ are essentially determined by
Eqs.~(\ref{xequi}) and (\ref{w}). Scaling, length to $\delta$ and
energy to $\hbar^2/m\delta^2=F\delta$, these two equations reduce,
in the zeroth order of $\xi/b$, to: $x_{\rm
eq}=(\delta/\xi)^{2}\delta$ and
$\omega/(F/m\delta)=(\xi/\delta)^4\ln{(b/\xi)}$, where we have
used $\hbar^2/2m\xi^2=\gamma$. Thus, to obtain the amplitude in
units of $\delta$ and the frequency in units of $F/m\delta$, we
need only to specify $\xi/\delta$ and $b/\delta$. The parameter
$b/\delta$ is a free fitting parameter that will be set a value to
get the best agreement with the numerical solution. In
Fig.~\ref{fig3}, we plot $x_0(t)$ obtained from the numerical
solution of the time-dependent Gross-Pitaevskii equation together
with the solution of Eq.~(\ref{eqnofm4}) using the same initial
conditions. The vortex is initially located at $x_0=-3.1\delta$
from the Thomas-Fermi surface and then evolved starting from rest.
The best agreement between the variational solution and the
numerical one was obtained for $b\simeq5\delta$. The coherence
length associated with this figure is calculated by fitting the
numerical density profile to Eq.~(\ref{chi2}), using $\xi$ as a
fitting parameter. This was done for both the $x$- and $y$-cross
sections of the vortex density profile as shown in
Fig.~\ref{fig4}. The average value of $\xi$ turns out to be
$\xi\simeq0.42\delta$. This is consistent with the Thomas-Fermi
estimate $\xi=2m\gamma/\hbar^2\simeq0.4\delta$, were we have used
$\gamma=gn(x_0)=-Fx_0$. Since the coherence length obtained using
the Thomas-Fermi approximation is approximately equal to the one
obtained numerically, we conclude that during the whole time
interval of Figs.~\ref{fig3} and \ref{fig4}, the vortex is moving
in a region where the Thomas-Fermi approximation is valid. At
first instance, we may take the average value of the coherence
length, namely $\xi=0.4\delta$, and then substitute it in $x_{\rm
eq}$ and $\omega$. This will lead to $x_{\rm eq}\simeq-6\delta$.
However, Fig.~\ref{fig3} indicates that the vortex oscillates
around $x_{\rm eq}\simeq-2.8\delta$. This discrepancy originates
probably from the difference in geometry of the system in the
variational calculation and the one in numerical calculation. In
the variational calculation, the vortex is moving in an infinite
background, but the energy integrations are performed within a
cylinder of radius $b=5\delta$. Furthermore, the number of atoms
within the cylinder is constant. In the numerical calculation, the
system is a rectangular grid of dimensions
$8\delta\times40\delta$. The number of atoms within the whole grid
is constant. Therefore, while in the variational calculation the
number of atoms is kept constant within an area of $\pi
b^2\simeq80\delta^2$, in the numerical calculation the number is
kept constant within an area of $320\delta^2$. Thus, there is a
difference in the density which leads to a difference in the
coherence length. In other words, the coherence length values
shown in Fig.~\ref{fig4} are not exactly those which we should use
in the variational calculation to seek agreement between the
results of the two calculations. We found, as shown in
Fig.~\ref{fig3}, that the best agreement with the numerical values
is obtained using $\xi=0.8\delta$.

We show also in Fig~\ref{fig3}, vortex trajectories for initial
distances $x_0(0)=-4\delta$ and $x_0(0)=-5\delta$. There is a
transient behavior in the beginning before acquiring the
oscillatory behavior. For initial starting distances larger than
$7\delta$, the oscillations disappear and the vortex moves in a
straight path parallel to the surface as predicted by the
variational calculation.
%This indicates that, in the numerical
%calculation, the number of atoms within a radius of $b=5\delta$,
%is 1/4 of the total number. Since the coherence length is
%inversely proportional to the density, we conclude that the
%coherence length is reduced by a factor of 4 in the numerical
%calculation in comparison to the variational one.

\section{Numerical solution}
\label{numsec} In this section, we describe our numerical
procedure in solving the time-dependent Gross-Pitaevskii equation.
Scaling length to $\delta$, energy to $\hbar^2/m\delta^2$, and
time to $m\delta^2/\hbar$, Eq.~(\ref{gp1}) takes the following
dimensionless form
\begin{equation}
\left[-{1\over2}\bmath{\nabla}^2_{\bf
\tilde{r}}+\tilde{x}+|\tilde{\psi}({\bf
\tilde{r}},\tilde{t})|^2\right]\tilde{\psi}({\bf
\tilde{r}},\tilde{t})=i{\partial\over\partial
\tilde{t}}\tilde{\psi}({\bf \tilde{r}},\tilde{t})\,,
\label{gp_scaled}
\end{equation}
where the tilde denotes scaled quantities and $|\psi|^2$ is scaled
to $1/4\pi a\delta^2$. It is interesting to notice here that the
scattering length disappears as a result of the scaling we employ
here.

We start by solving the time-independent Gross-Pitaevskii equation
 \begin{equation}
\left[-{1\over2}{\partial^2\over\partial
\tilde{x}^2}+\tilde{x}+|\tilde{\psi}_0(
\tilde{x})|^2\right]\tilde{\psi}_0( \tilde{x})=0\,, \label{gpind}
\end{equation}
with a vortex-free initial state. We use the imaginary-time
evolution method to solve this equation. The resulting density
$\tilde{n}_0(\tilde{x})$ is the background density profile for the
vortex. We multiply this background density by a density profile
of a vortex in a uniform background, namely $\chi^2$ given by
Eq.~(\ref{chi2}). The resulting density profile can not yet be
used as an initial state for solving the time-dependent
Gross-Pitaevskii equation. This is so since $\chi$ of
Eq.~(\ref{chi2}) is not the exact density profile of the vortex.
When we simulated the dynamics of the vortex using this density
profile, the core of the vortex acquired breathing oscillations
during the vortex motion. These oscillations produced circular
density waves emitted out of the vortex core through the
background and then reflected from the boundaries to interfere
with vortex affecting its trajectory. Since we are interested only
in the center-of-mass motion of the vortex, it is necessary to use
the exact density profile of the vortex to avoid exciting such
density waves. This can be obtained by evolving in imaginary time
for a certain time period the density $\tilde{n}_0\chi^2$. In
other words, we resolve Eq.~(\ref{gpind}) with
$\sqrt{\tilde{n}_0\chi^2}$ as an initial wavefunction. The
resulting profile contains the exact background and vortex core
structure. This profile is then used as the initial profile for
the time-dependent Gross-Pitaevskii equation. We found that
density waves emerging from the vortex core have now almost
disappeared. However, evolving in imaginary time results in the
vortex sliding towards the surface of the condensate and
ultimately disappearing there. In the present procedure, we evolve
for only a finite time interval long enough to ensure adequate
suppress of the density waves before the vortex reaches the
surface. The distance from the surface reached by the vortex at
the end of this time interval is to be considered as the initial
distance for the real time evolution. We checked that our results
for the real-time dynamics of the vortex are not affected by the
length of this time interval once the density waves are
significantly suppressed.

We employed a fourth-order Runge-Kutta method to integrate the
time derivative of the time-dependent Gross-Pitaevskii equation.
The spacial grids we used are $8\delta\times40\delta$ and
$8\delta\times80\delta$ with spacial descretization of $\Delta
x=0.1\delta$ and time descretization $\Delta {\tilde t}=10^{-5}$.
The long side of the grids is along the $y$-axis since the vortex
is faster in that direction. We have used the boundary conditions
that the derivative of density with respect to $y$ vanishes at the
two edges of the grid normal to the $y$-axis. The other two edges
were left free. We have used larger grids to check that the
boundaries do not affect the results obtained here and found that
the above-mentioned grid sizes are adequate for that purpose. We
have also calculated the total number of atoms and energy to find
that they are fluctuating within less than $0.1\%$. The
fluctuations were random and not correlated to the vortex motion.

\begin{figure}[htb]
\begin{center}
\includegraphics[width=10.cm]{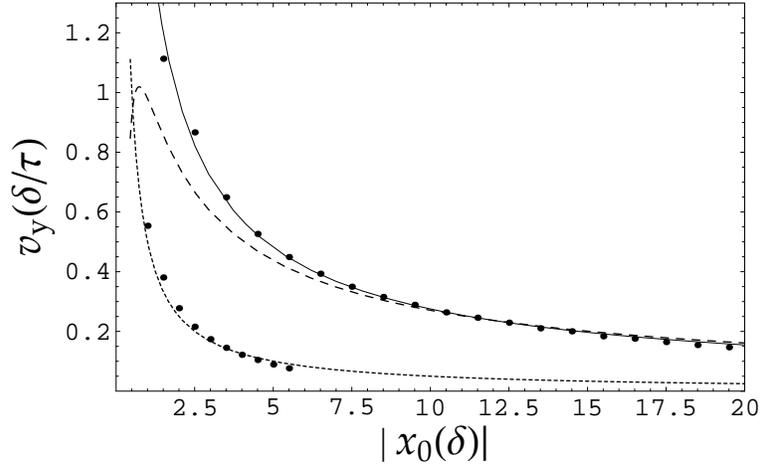}
\end{center}
\caption[a]{ The vortex velocity in the $y$-direction as a
function of the magnitude of the distance between the vortex core
and the Thomas-Fermi surface. The velocity is in units of
$\delta/\tau$. The surface depth $\delta$ is defined by
$\hbar^2/m\delta^2=F\delta$, and the time $\tau$ is defined by
$\tau=\hbar/F\delta$. The dots on the upper curve are the results
of the numerical solution of a vortex moving near the surface of
the condensate. The dots on the lower curve correspond to a vortex
moving parallel to a hard wall in a uniform density background.
The solid curve is the result of the variational calculation of
section~\ref{sec1}, given by Eq.~(\ref{vy2}). The long-dashed
curve corresponds to the expression derived by
Anglin~\cite{james2}. The short-dashed curve is given by
$v_y=\hbar/2mx_0$.} \label{fig2}
\end{figure}

\begin{figure}
\begin{center}
\includegraphics[width=10.cm]{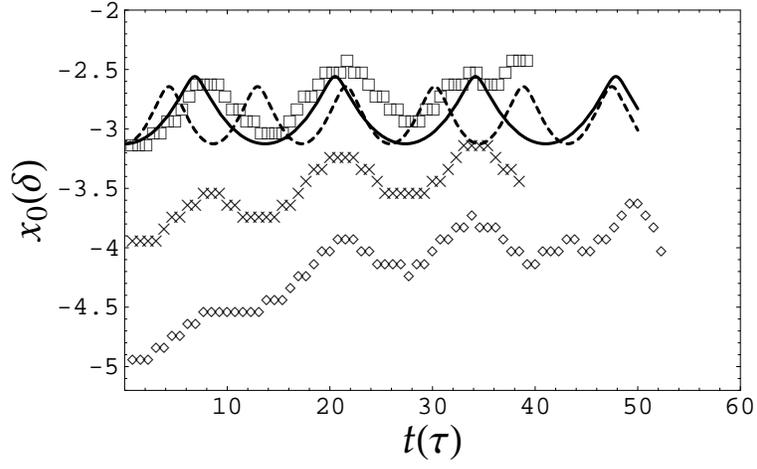}
\end{center}
\caption[a]{ The $x$-coordinate of the vortex core measured from
the Thomas-Fermi surface. The points are the results of the
numerical solution of the time-dependent Gross-Pitaevskii
equation, Eq.~(\ref{gp1}) for different initial distances. The
initial velocity is taken to be zero. The solid and dashed curves
are the results of the variational calculation of
subsection~\ref{sub3}. The solid curve is the solution of
Eq.~(\ref{eqnofm4}), and the dashed curve is an analytic
approximate solution given by Eq.~(\ref{sola}). The value of $b$
used here is $5\delta$. } \label{fig3}
\end{figure}

\begin{figure}[htb]
\begin{center}
\includegraphics[width=10.cm]{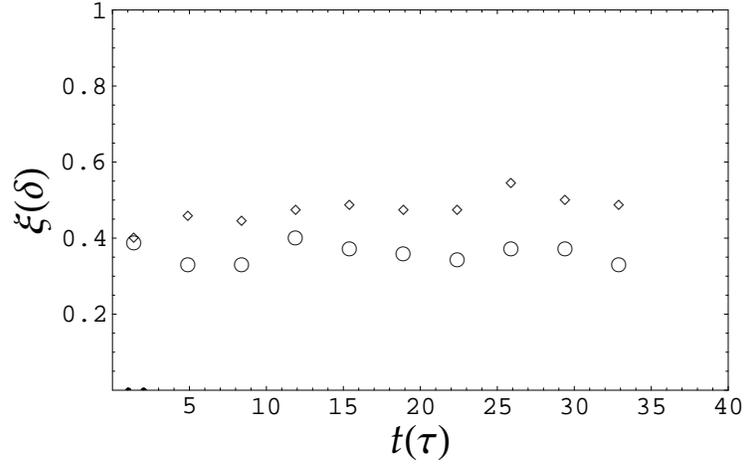}
\end{center}
\caption[a]{The coherence length associated with the $x$- and
$y$-cross sections of the vortex density profile. It is calculated
by fitting the numerical density profile of the vortex to
Eq.~(\ref{chi2}). Circles represent the coherence length
associated with the $y$-cross section of the vortex density
profile, and diamonds represent the coherence length for the
$x$-cross section. } \label{fig4}
\end{figure}

To test the accuracy of our numerical calculation, we start by
simulating the dynamics of a vortex near a hard wall in a
condensate of uniform background density. The vortex velocity is
calculated theoretically to be $v_y=\hbar/2mx_0$ \cite{james2}. In
Fig.~\ref{fig2}, we show that our code accurately generates such a
dependence for the vortex velocity.

\begin{figure}[htb]
\begin{center}
\includegraphics[width=15.cm]{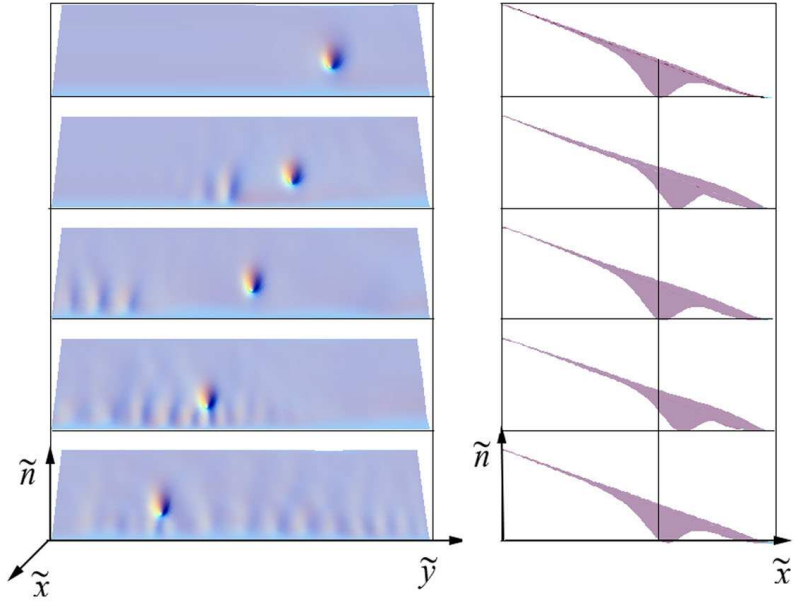}
\end{center}
\caption[a]{Density profile of the vortex. The upper row of
pictures corresponds to the initial state with the vortex core
located at $x_0(0)=-3.1\delta$. The left column of pictures is a
front view in which the $x$-axis is normal to the picture towards
the viewer. The right column is a side view in which the $y$-axis
is normal to the picture towards the viewer. The vertical line in
the right column of pictures is to reference the core location
with respect to the initial position, and shows clearly that the
vortex core oscillates in the $x$-direction.} \label{fig5}
\end{figure}

\begin{figure}[htb]
\begin{center}
\includegraphics[width=15.cm]{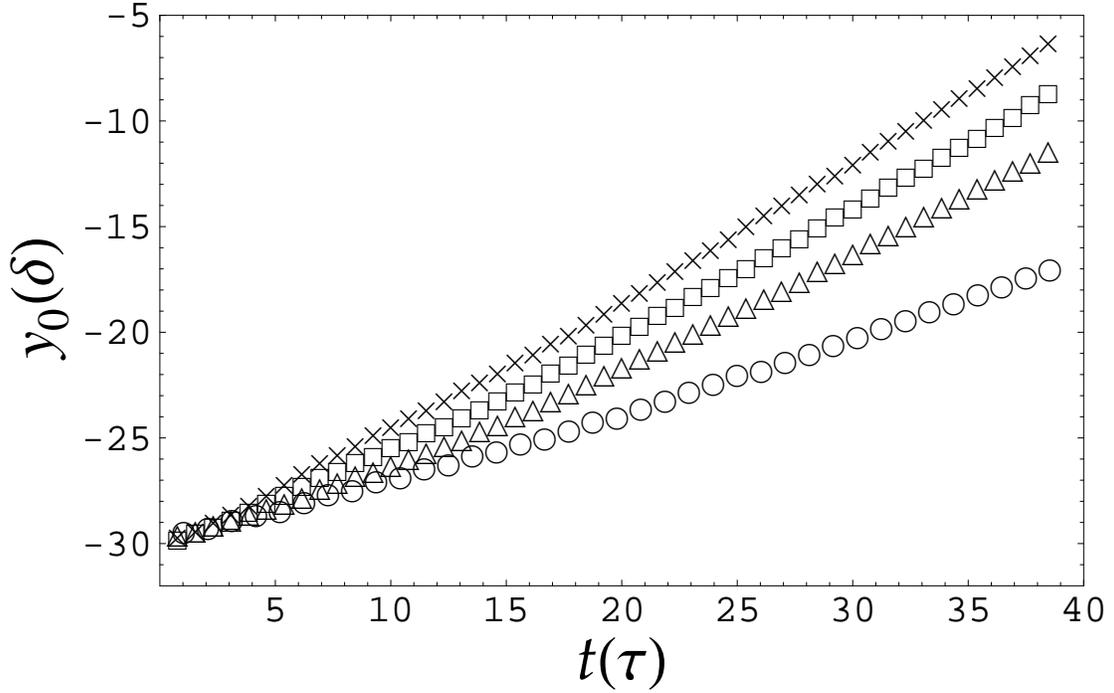}
\end{center}
\caption[a]{ The vortex $y$-coordinate. Starting from the upper
curve, the values of the initial position of the vortex core $x_0$
are: $-3\delta$, $-4\delta$, $-5\delta$, and $-7\delta$. The
constant slope indicates that the velocity component parallel to
the surface is constant.} \label{fig6}
\end{figure}

Having checked the accuracy of our numerical procedure, we
simulate the dynamics of the vortex in the surface region of the
condensate. In Fig.~\ref{fig5}, we show the results in the form of
a number of snapshots of the density of the condensate visualizing
the realtime motion of the vortex core along the surface of the
condensate. In Figs.~\ref{fig3} and \ref{fig6}, we plot the vortex
coordinates $x_0$ and $y_0$ versus time. These two figures show
that while the vortex is moving with a constant velocity in the
$y$-direction, it oscillates in the $x$-direction as predicted in
the previous section. To investigate the $x_0$-dependance of
$v_y$, we calculate the numerical value of $v_y$ for a number of
different values of $x_0$. The results are shown in
Fig.~\ref{fig2} where we also plot the theoretical prediction of
Anglin \cite{james2} and the result of our variational
calculation. The formula derived by Anglin is supposed to be valid
away from the Thomas-Fermi surface. Figure~\ref{fig2} shows indeed
that while this formula agrees well with the numerical calculation
for large $x_0$, it deviates from it for values of $x_0$ of order
$\delta$. On the other hand, this figure shows that our formula,
derived in subsection~\ref{sub2}, is in a better agreement with
the numerical data especially for $x_0<10\delta$ where the formula
of Anglin starts to deviate from the numerical values.

We have also calculated the coherence length along both the $x$
and $y$-directions, which is of the order of the size of the core
of the vortex. This is plotted in Fig.~\ref{fig4}. This figure
shows that the core sizes change slightly over the whole time
interval considered. The average value of $\xi$ in this figure
agrees with the Thomas-Fermi estimate as explained in
subsection~\ref{sub3}. In the previous section, we performed the
calculation using the Thomas-Fermi approximation for the
background density and neglected the dynamics of the size of the
core. It is clear now that this is indeed justified here.

\section{Conclusions}
\label{conc} We have shown that our variational calculation
describes at least qualitatively the motion of a vortex near the
surface of a Bose-Einstein condensate, and generates results that
agree with previous calculations \cite{james1,fetter2}. These
qualitative results were then supported and quantified using a
numerical simulation of the vortex motion.

One of the main findings of this work is Eq.~(\ref{vy2}) which
gives the $y$-component (parallel to the surface) of the vortex
velocity as a function of it is distance from the Thomas-Fermi
surface $x_0$. It agrees qualitatively with that of
Anglin~\cite{james2}, and describes more accurately $v_y$ for
values of $x_0$ of order of $\delta$ as shown in Fig.~\ref{fig2}.
This figure shows clearly that Eq.~(\ref{vy2}) can be used for
both small and large $x_0/\delta$, unlike the result of
Anglin~\cite{james2} which is accurate only for large
$x_0/\delta$.

Another main result of this work is that, if the vortex is moving
within few coherence lengths $\xi$ from the Thomas-Fermi surface,
it will have an oscillatory component of its velocity in the
direction normal to the surface. These oscillations disappear for
$x_0\ll-\delta$ and the vortex will have only the $y$-component of
the velocity. To further understand the physics of these
oscillations, we have shown in subsection~\ref{sub2} that the
vortex {\it feels} an effective potential that is a function of
$x_0$ and has local minimum for negative values of $x_0$. This
effective potential is composed mainly of a sum of a
mean-field-energy and a trapping potential energy part as shown in
Eq.~(\ref{defunc4}). These two parts compete such that the
effective potential has a minimum at $x_{\rm
eq}=-(\delta/\xi)^{2}\delta$. Using a collective coordinates
approach, we have shown that the vortex oscillates around this
minimum with a frequency
$\omega=\sqrt{(F^4\xi^2/8m\gamma^3)\ln{(b/\xi)}}$. As discussed in
subsection~\ref{sub2}, this frequency is of order less than
$\sqrt{|x_0|^{-4}\ln{(|x_0|/\xi)}}$. Both this frequency and the
$x_0$-dependent part of $U(x_0)$ vanish for large $x_0$. This
means that the vortex oscillations will disappear away from the
surface of the condensate and will be noticeable only when
$\xi/x_0$ is not negligible. To the best of our knowledge, these
oscillations are predicted here for the first time. As a possible
alternative explanation to these vortex oscillations near the
surface, we mention the possibility that the vortex being dragged
by a collective or surface wave. In our numerical procedure, the
evolution in imaginary time which generates the initial state for
the real time evolution, may include some excitations of
collective modes. Further investigation is needed to verify this
argument.

It was shown in Ref.~\cite{james2} that the energy of a vortex
near the Thomas-Fermi surface is not infrared divergent as in the
bulk. This is so since the vortex is localized to within its
distance from the surface. To avoid these infrared divergencies in
the energy, we have introduced an infrared cut-off on the energy.
This is equivalent to using the infrared-free flow pattern of
Ref.~\cite{james2} which gives finite energy.

\appendix
\section{Details of the calculation of the lagrangian}
\label{app} In this Appendix we present the details of calculating
the lagrangian Eq.~(\ref{lagapp2}). This is performed by
substituting the trial trial wave function
Eq.~(\ref{psitrial_norm}) in Eq.~(\ref{lagrangian}).

We start by calculating the energy functional Eq.~(\ref{energy}).
According to the geometry illustrated in Fig.~\ref{fig1}, the
cartesian coordinates $x^\prime=x-x_0$ and $y^\prime=y-y_0$ with
origin at the core of the vortex are related to the polar
coordinates $\rho(x^\prime,y^\prime)$ and
$\phi(x^\prime,y^\prime)$ by
$\rho(x^\prime,y^\prime)=\sqrt{{x^\prime}^2+{y^\prime}^2}$ and
$\phi(x^\prime,y^\prime)=\tan^{-1}(y^\prime/x^\prime)$. This leads
to that $\partial\phi/\partial x=\partial\phi/\partial
x^\prime=\cos\phi/\rho$ and $\partial\phi/\partial
y=\partial\phi/\partial y^\prime=-\sin\phi/\rho$. In addition, we
have $\partial\rho/\partial x=\partial\rho/\partial
x^\prime=\sin\phi$ and $\partial\rho/\partial
y=\partial\rho/\partial y^\prime=\cos\phi$. The derivative of the
trial wave function $\psi(x,y,t)$ with respect to $x$, can thus be
shown to take the form
\begin{eqnarray}
{\partial\psi\over\partial x}&=& \sqrt{N\over
lN_1}\exp{\left[i(\Phi+(m/\hbar){\bf v\cdot r
})\right]}\left[{\chi\over2\sqrt{xx_0}}+\sqrt{x\over
x_0}\chi_\rho\sin{\phi}\right.\nonumber\\
&+&\left.i\sqrt{x\over x_0}\chi\left({\cos{\phi}\over\rho}
+{1\over2x_0}\cos{\phi}\sin{\phi}+{m\over\hbar}v_x\right)\right]
\label{psix}~.
\end{eqnarray}
The contribution to the kinetic energy is obtained by integrating
$|\partial\psi/\partial x|^2$ with respect to $\rho$ and $\phi$.
The latter integral reduces to
\begin{eqnarray}
\int_0^{2\pi}d\phi\,\left|{\partial\psi\over\partial
x}\right|^2&=& 2\pi{N\over
lN_1}\left[{\chi^2\over4\sqrt{x_0^4-x_0^2\rho^2}}+{1\over2}\chi_\rho^2\right.\nonumber\\
&+&\left.\chi^2\left({1\over2\rho^2}+{m^2v_x^2\over\hbar^2}+{1\over8x_0^2}\right)\right]
\label{psix2}~,
\end{eqnarray}
where $\chi_\rho$ denotes $d\chi(\rho)/d\rho$. It should be noted
that we neglected here the contributions of order
$(\rho/x_0)^2(\ln{(\alpha\rho/x_0))^2}$ that originate from the
logarithmic term of the phase $\Phi(\rho,\phi)$. In deriving this
term, Anglin \cite{james2} performed a perturbative expansion of
the phase of the vortex near the core using $\xi/x_0$ as the
perturbation parameter, where $\xi$ is the coherence length
defined in subsection \ref{sub1}. Since in Anglin's calculation
only terms of order $(\xi/\rho)\ln{(\alpha\rho/x_0)}$ are kept, it
will be inconsistent if we keep here higher order terms. We keep,
however, the higher order terms that do not originate from the
logarithmic term of the phase.

Similarly, for the derivative with respect to $y$, we have
\begin{eqnarray}
{\partial\psi\over\partial y}&=& \sqrt{N\over
lN_1}\exp{\left[i(\Phi+(m/\hbar){\bf v\cdot r
})\right]}\left[\sqrt{x\over
x_0}\chi_\rho\cos{\phi}\right.\nonumber\\
&+&\left.i\sqrt{x\over x_0}\chi\left({-\sin{\phi}\over\rho}
+{1\over2x_0}\ln{\alpha\rho\over
x_0}+{1\over2x_0}\cos{\phi}^2+{m\over\hbar}v_y\right)\right]
\label{psiy}~,
\end{eqnarray}
which leads to
\begin{eqnarray}
\int_0^{2\pi}d\phi\,\left|{\partial\psi\over\partial
y}\right|^2&=& 2\pi{N\over
lN_1}\left[{1\over2}\chi_\rho^2+\chi^2\left({1\over2\rho^2}+{m^2\over\hbar^2}v_y^2\right.\right.\nonumber\\
&-&\left.\left.{m\over\hbar x_0 }v_y({1\over2}-\ln{\alpha\rho\over
x_0})-{1\over2x_0^2}({1\over4}+\ln{\alpha\rho\over
x_0})\right)\right] \label{psiy2}~.
\end{eqnarray}
The kinetic energy is then given by
\begin{eqnarray}
KE&=&{\hbar^2\over2m}\int_0^ldz\int_0^bd\rho\rho\int_0^{2\pi}d\phi\left|\bmath{\nabla}\psi({\bf
r})\right|^2\nonumber\\ &=&{\hbar^2\over2m}{2\pi N\over
N_1}\int_0^bd\rho\rho\left[{\chi^2\over4x_0^2\sqrt{1-\rho^2/x_0^2}}+\chi_\rho^2\right.\nonumber\\
&+&\left.\chi^2\left({1\over\rho^2}+{m^2\over\hbar^2}(v_x^2+v_y^2)-{m\over2\hbar
x_0 }v_y(1-2\ln{\alpha\rho\over x_0})-{1\over
2x_0^2}\ln{\alpha\rho\over x_0}\right) \right]
%\nonumber\\
%&=&{\hbar^2\over2m}2\pi N
%\left[{N_2(x_0)+N_3+N_4\over N_1}+{m^2\over\hbar^2}(v_x^2+v_y^2)\right.\nonumber\\
%&-&\left.{m\over\hbar x_0 }v_y({1\over2}-{N_5\over N_1})-{1\over
%2x_0^2}{N_5\over N_1} \right]
\label{keapp}~.
\end{eqnarray}

The potential energy associated with the external force is
\begin{eqnarray}
PE&=&\int_0^ldz\int_0^bd\rho\rho\int_0^{2\pi}d\phi\,
Fx\left|\psi({\bf r })\right|^2\nonumber\\
&=&F{2\pi N\over N_1}\int_0^bd\rho\rho
\chi^2\left(x_0+{\rho^2\over2x_0}\right) \label{peapp}~.
\end{eqnarray}

The mean-field energy is
\begin{eqnarray}
MFE&=&\int_0^ldz\int_0^bd\rho\rho\int_0^{2\pi}d\phi\,
{1\over2}g\left|\psi({\bf r })\right|^4\nonumber\\
&=&{1\over2}g{2\pi N\over N_1}\int_0^bd\rho\rho
\chi^4\left(1+{\rho^2\over2x_0^2}\right) \label{mfeapp}~.
\end{eqnarray}

Combining these three contributions, we find the energy functional
per atom
\begin{eqnarray}
E[x_0,v_x,v_y]/N&=&
{\hbar^2\over2m}(N_2(x_0)+N_3)+{1\over2}m(v_x^2+v_y^2)-{\hbar\over4
x_0 }v_y(1-2N_4(x_0))\nonumber\\
&-&{\hbar^2N_4(x_0)\over4mx_0^2}+F\left(x_0+{N_5\over2x_0}\right)
+{1\over2}\gamma\left(N_6+{N_7\over2x_0^2}\right)
\label{efapp}~,
\end{eqnarray}
where $\gamma=gN/N_1l$ and we have defined the coefficients
\\
\begin{math}
N_2(x_0)={2\pi\over N_1
}\int_0^bd\rho\rho{\chi^2/4x_0^2\sqrt{1-\rho^2/x_0^2}}~,
\end{math}
\\
\begin{math}
N_3={2\pi\over N_1
}\int_0^bd\rho\rho\left(\chi_\rho^2+{\chi^2/\rho^2}\right)~,
\end{math}
\\
\begin{math}
N_4(x_0)={2\pi\over N_1 }\int_0^bd\rho\rho\chi^2\ln{(\alpha\rho/
x_0)}~,
\end{math}
\\
\begin{math}
N_5={2\pi\over N_1 }\int_0^bd\rho\rho^3\chi^2~,
\end{math}
\\
\begin{math}
N_6={2\pi\over N_1 }\int_0^bd\rho\rho\chi^4~,
\end{math}
\\
\begin{math}
N_7={2\pi\over N_1 }\int_0^bd\rho\rho^3\chi^4~.
\end{math}

The relevant energy is the energy change due to the presence of
the vortex. Therefore we have to subtract from the above energy
functional the energy of a vortex-free background
\begin{eqnarray}
E_{\rm vf}&=& \int_0^bd\rho\rho\int_0^{2\pi}d\phi\left[ {1\over
2}g\left({Nx\over\pi b^2 lx_0}\right)^2 +F\left({Nx\over \pi
b^2lx_0}\right)x \right]\label{evfapp}\,.
\end{eqnarray}
Thus, the energy difference associated with vortex reads
\begin{eqnarray}
\Delta E[\psi,\psi^*]/N&=&
(N_2(x_0)+N_3){\hbar^2\over2m}+{1\over2}m(v_x^2+v_y^2)-(1-2N_4(x_0)){\hbar\over4
x_0 }v_y\nonumber\\
&-&{\hbar^2N_4(x_0)\over4mx_0^2} +F\left({N_5\over2}-{
b^2\over4}\right){1\over x_0}\nonumber\\
&+&{1\over2}\gamma\left[N_6-{N_1\over\pi
b^2}+\left({N_7\over2}-{N_1\over4\pi}\right){1\over
x_0^2}\right]\label{defunc2app}~.
\end{eqnarray}
The time derivative of the trial wave function is given by
\begin{eqnarray}
{\partial\psi\over\partial t}&=& \sqrt{N\over
lN_1}\exp{\left[i(\Phi+(m/\hbar){\bf v\cdot r
})\right]}\sqrt{x\over x_0 }\left[-{{\dot x }_0\over
2x_0}\chi+\chi_\rho{\partial\rho\over\partial t
}+i\chi\left({\partial\Phi\over\partial t}+{m\over\hbar}({\dot
v}_xx+{\dot v}_yy)\right)\right]\label{tdapp}~.
\end{eqnarray}
Using that $\partial\rho/\partial t=(\partial\rho/\partial
x^\prime)(\partial x^\prime/\partial t)+(\partial\rho/\partial
y^\prime)(\partial y^\prime/\partial
t)=-{{\dot{x}_0^\prime}}\sin{\phi}-{{\dot{y}_0^\prime}}\cos{\phi}$
and similarly $\partial\phi/\partial
t=(-{{\dot{x}_0^\prime}}\cos{\phi}+{{\dot{y}_0^\prime}}\sin{\phi})/\rho$
and Eq.~(\ref{jphase}), the part of the lagrangian (per atom)
containing the time derivatives takes the form
\begin{eqnarray}
&\,&{i\hbar\over2N}\int_0^ldz\int_0^bd\rho\rho\int_0^{2\pi}d\phi
\left(\psi^*\dot{\psi}-\psi\dot{\psi}^*\right)=\nonumber\\
&-&\hbar{\dot{y}_0\over
4x_0}(1-2N_4(x_0))-m(x_0\dot{v}_x+y_0\dot{v}_y)-N_5{m\dot{v}_x\over2x_0}
\label{tlapp}~.
\end{eqnarray}
The lagrangian per atom, Eq.~(\ref{lagapp2}), is then obtained.
%\begin{eqnarray}
%L/N&=&-(1-2N_4(x_0))\hbar{\dot{y}_0\over
%4x_0}-m(x_0\dot{v}_x+y_0\dot{v}_y)-N_5{m\dot{v}_x\over2x_0}\nonumber\\
%&-&{\hbar^2\over2m}(N_2(x_0)+N_3)-{1\over2}m(v_x^2+v_y^2)+(1-2N_4(x_0)){\hbar\over4
%x_0 }v_y\nonumber\\
%&+&{\hbar^2N_4(x_0)\over4mx_0^2} -F\left({N_5\over2}-{
%b^2\over4}\right){1\over x_0}\nonumber\\
%&-&{1\over2}\gamma\left[N_6-{N_1\over\pi
%b^2}+\left({N_7\over2}-{N_1\over4\pi}\right){1\over
%x_0^2}\right]\label{lagapp2app}~.
%\end{eqnarray}

\section*{Acknowledgments}

I would like to thank Halvor Nilson from NORDITA for useful
discussions. I would also like  to thank the  anonymous referee
whose comments have significantly enhanced this paper.

%\end{multicols}

\end{document}